\documentclass[aps,prb,twocolumn,showpacs,superscriptaddress]{revtex4-1}
\usepackage{amsmath}
\usepackage{amssymb}
\usepackage[dvips]{graphicx}
\usepackage{dcolumn}
\usepackage{color}
\begin{document}

\title{Voltage control of the spin-dependent interaction constants of Dipolaritons and its application to Optical Parametric Oscillator}

\author{A. V. Nalitov}
\affiliation{Institut Pascal, PHOTON-N2, Clermont Universit\'{e}, Blaise Pascal
University, CNRS, 24 avenue des Landais, 63177 Aubi\`{e}re Cedex, France.}
\author{D. D. Solnyshkov}
\affiliation{Institut Pascal, PHOTON-N2, Clermont Universit\'{e}, Blaise Pascal
University, CNRS, 24 avenue des Landais, 63177 Aubi\`{e}re Cedex, France.}
\author{N. A. Gippius}
\affiliation{Skolkovo Institute of Science and Technology, 100 Novaya st., Skolkovo, Odintsovsky district, Moscow Region, Russia 143025}
\affiliation{A. M. Prokhorov General Physics Institute, RAS, Vsavilova Street 38, Moscow 119991, Russia}
\author{G. Malpuech}
\affiliation{Institut Pascal, PHOTON-N2, Clermont Universit\'{e}, Blaise Pascal
University, CNRS, 24 avenue des Landais, 63177 Aubi\`{e}re Cedex, France.}

\begin{abstract}
Dipolariton is a voltage controlled mixture of direct and indirect excitons in asymmetric double quantum wells coupled by resonant carrier tunneling, and a microcavity photon \citep{Cristofolini2012}.
We calculate the voltage dependence of the spin-singlet and spin-triplet interaction parameters $\alpha_{1}$ and $\alpha_{2}$. Both parameters can reach values one order of magnitude larger than that of exciton-polaritons thanks to the strong interaction between indirect excitons.
We show that the variation of the indirect exciton fraction of the dipolaritons induces a  changes of sign of $\alpha_{2}$ increasing voltage: the interaction passes from attractive to repulsive. For large enough voltage $\alpha_{2}$  becomes larger than $\alpha_{1}$ which in principle can lead to the formation of a circularly polarised dipolariton condensate. We propose an application of the $\alpha_{2}$ dependence to a voltage-controlled dipolaritonic optical parametric oscillator. The change of sign of $\alpha_{2}$  allows an on-site control of the linear polarization degree of the optical signal and its on-demand inversion.
\end{abstract}

\maketitle

\section{I. Introduction.}


%
Cavity exciton-polaritons are mixed exciton-photon quasiparticles formed by the strong coupling between cavity photons, and quantum well excitons \cite{Microcavities2007}. They interact strongly between each other because of their excitonic component. They represent one of the best examples of interacting photons implementing the concept of photonic quantum fluid \cite{CarusottoRMP}. From an applied point of view, these strongly interacting photonic particles represent a unique opportunity for the realization of low threshold nonlinear optical devices \cite{Savvidis2000,Messin2001,Paraiso2010,Gao2012,Nguyen2013,Sturm2013}. The typical way to modify the strength of the polariton-polariton interaction is to change the excitonic content of the polariton by changing the energy detuning between the bare exciton and photon energy \cite{Feng2013}. Using this approach, changing the interaction constants means changing the position of the experiment on a wedged sample. Polaritons are also spinor particles with a spin structure similar to the one of photon \cite{Shelykh2010}. Their interactions are strongly spin-anisotropic. Indeed, quantum well 1s-excitons do not possess a dipole moment, and the main mechanism of their interaction is the short range exchange interaction \cite{Ciuti1998}. We will call the interaction parameter in the triplet configuration (same spins) $\alpha_1$. If one considers two polaritons with opposite spins, the exchange of the carriers of their excitonic part leads to the formation of dark excitonic states of total angular momentum $\pm 2$, whose separation from the polariton states is of the order of the Rabi splitting. An alternative mechanism of interaction between dipolaritons having opposite spins is associated with the formation of an intermediate biexciton state \cite{Glazov2009,Vladimirova2010,Takemura2014}. As a result, interaction between polaritons of opposite spins described by an interaction parameter $\alpha_2$ is a second order process. One should notice that far from the biexciton resonance it is strongly suppressed compared to the first order carrier exchange interaction $\alpha_1$, and is weakly attractive. This fact had numerous consequences, verified by multiple experiments, such as the linear polarisation of polariton condensates \cite{Shelykh2006,Kasprzak2006,Kasprzak2007}, structure of topological defects \cite{Rubo2007, Solnyshkov2012}, polarization inversion in polariton optical parametric oscillators \cite{Renucci2005}, multistability effects \cite{Gippius2007,Paraiso2010} and others.  On the other hand when the polariton state is approaching the bi-exciton energy, $\alpha_2$ increases and it can even change sign while crossing the resonance \cite{Vladimirova2010,Takemura2014}. This interesting mechanism of control of the sign of the interaction parameter is however necessarily associated with strong losses in the resonant bi-exciton state, and also by a large thermal depletion of the polariton states \cite{Vishnevsky2012, Paraiso2010}.

If we now consider the indirect excitons (IX) in coupled quantum wells (CQWs), they are formed by an electron and a heavy hole in neighboring CQWs and thus have a dipole moment oriented along the growth direction (due to the applied bias) and proportional to the CQWs separation distance $d$.
Consequently, the dipole-dipole repulsion of IXs is a first-order effect and is even stronger than their exchange interaction, which switches from repulsive to attractive while increasing $d$ \cite{Kyriienko2012,Nalitov2013}.
However, the coupling of an IX with a cavity photon mode is limited by the small oscillator strength of the IX, proportional to the overlap of the vanishing tails of the electron and hole wavefunctions  in the barriers.

\begin{figure} \label{fig1}
\includegraphics[width=0.99\linewidth]{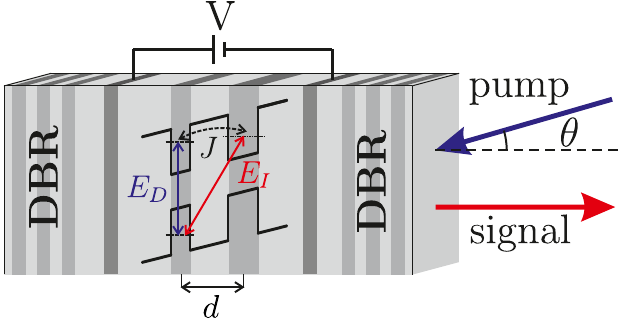}
\caption{A sketch of the studied system: asymmetric double quantum well (ASDQWs) is situated between the contacts and two Bragg reflectors forming a microcavity. 
Voltage $V$, applied to the contacts, shifts the energy diagram of the ASDQW and the energies of the indirect ($E_{\mathrm{I}}$) and direct ($E_{\mathrm{D}}$) exciton states, coupled via tunneling of the electron through the barrier.}
\end{figure}

Recently, exploiting the asymmetric double quantum wells (ASDQWs) for resonant tunnel coupling of IX to the conventional direct exciton (DX), their bound state was suggested and realized \cite{Cristofolini2012}.
The  ASDQW embedded in a microcavity structure is schematically shown on the Figure 1.
The coupling of three modes: indirect and direct exciton and cavity photon, gives rise to new two-dimensional quasi-particles \emph{dipolaritons}.
Being thus a mixture of dipolar matter and light, dipolaritons represent  photons with strong dipolar interaction. 

In this work we theoretically describe the voltage dependence of the spin anisotropic interaction between dipolaritons, and see how this dependence can be exploited in a practical application.
We demonstrate that the interaction of dipolaritons is at least one order of magnitude stronger than the one of conventional polaritons. 
We also show that the nature of the interaction between dipolaritons of opposite spins can be switched from attractive to repulsive. 
We illustrate this dependence by calculating the polarisation response of a dipolaritonic optical parametric oscillator (DOPO).
This effect manifests itself in the linear polarization inversion of the DOPO that can be switched by an applied voltage.
We also predict bistable behavior of the DOPO emission versus the pumping intensity and the applied voltage.

The present work is organized as follows.
The spin-dependent dipolariton wave functions are calculated in the section II.
The calculation of the matrix elements of the interaction between dipolaritons is described in Section III.
The analysis of the suggested dipolaritonic OPO scheme is given in Section IV.
Discussion of the obtained results concludes the work in Section V.

\section{General Description, Calculation of the dipolariton wavefunctions.}

\begin{figure}[t] \label{fig2}
\includegraphics[width=0.99\linewidth]{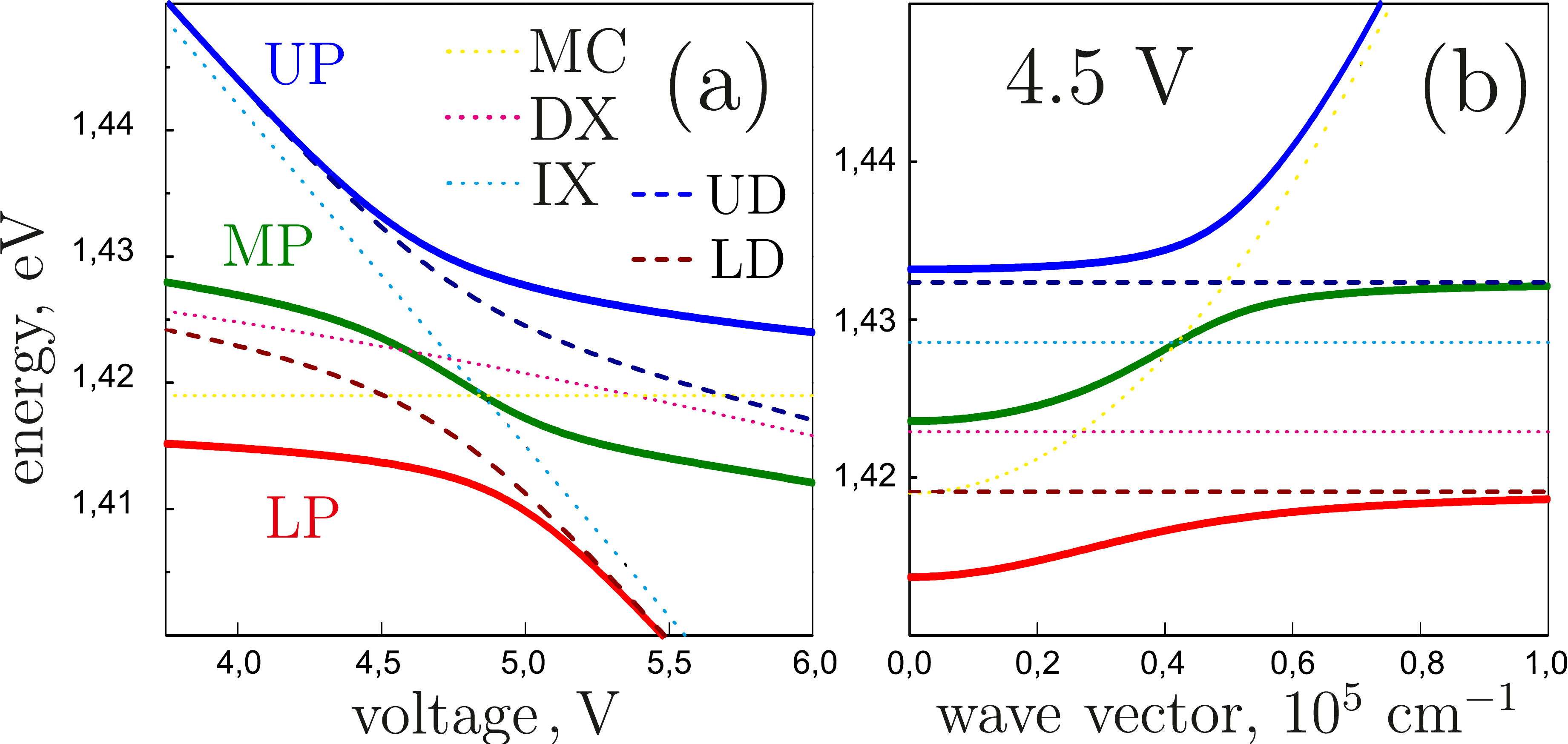}
\caption{ (color online) Dipolaritonic branches stemming from direct (DX), indirect (IX) excitonic and microcavity photonic modes. (a) Lower (LP), middle (MP) and upper (UP) dipolaritonic branches (solid lines), lower (LD) and upper (UD) dark exciton branches (dashed lines), uncoupled DX,IX and microcavity energies (dotted lines) versus applied voltage at zero wave vector. (b) The same branches versus wave vector at specific voltage (4.5 V). Parameters of the branches calculation are taken to fit the results of the reference \citep{Coulson2013}. }
\end{figure}

We consider a structure consisting of ASDQWs embedded in a microcavity composed by two Bragg mirrors (figure 1).
The DX state is formed by an electron-hole pair in the ground state, confined in one QW, while the IX consists of a hole in the same QW as the one of the DX, and an electron in the other QW.
The DX and IX are thus coupled via the electron tunneling through the barrier, described by the coupling constant $J$, while the DX coupling to the cavity mode is induced by the exciton oscillator strength, giving rise to the Rabi splitting $\Omega$.

The ASDWQs are subject to an external electric field, normal to their plane, produced by a voltage $V$, applied to the contacts on the doped layers \citep{Coulson2013}.
The field shifts electron and hole levels of size quantization in both QWs, so that the DX energy $E_\mathrm{D}(V)=E_\mathrm{D}(0)-\beta V^2$ slowly decreases, depending quadratically on the field, due to the quantum confined Stark effect \cite{Miller1984}. On the other hand, the IX energy $E_\mathrm{I}(V)=E_\mathrm{I}(0)-\gamma V$ shift is steeper and depends linearly on the field with the proportionality coefficient being the IX dipole moment\cite{Sivalertporn2012}.
In the absence of field, $E_{\mathrm{I}}(0) > E_{\mathrm{D}}(0)$. Therefore, at a certain voltage $V_0$ both exciton states have the same energy and become resonantly coupled.
We consider the range of voltages around $V_0$, where the IX energy shift is smaller than the energy distance to the nearest electron confinement level or the closest cavity photon mode.
This allows to neglect the presence of other states and to write the following system Hamiltonian \cite{Cristofolini2012}:
\begin{equation} \label{Hamiltonian}
H(Q,V) = \left( \begin{matrix}
E_{\mathrm{I}}(V) & -J/2 & 0 \\
-J/2 & E_{\mathrm{D}}(V) & -\Omega/2 \\
0 & -\Omega/2 &  E_{\mathrm{C}} + T_{\mathrm{C}}(Q)
\end{matrix} \right),
\end{equation}
where $T_{\mathrm{C}}(Q)=\hbar^2 Q^2/2 m_{\mathrm{C}}$ term accounts for the propagation of light in the cavity plane and represents the kinetic energy of the confined photon.
Here $\mathbf{Q}$ is the wave vector in the cavity plane and $m_{\mathrm{C}}$ stands for the effective mass of the cavity photon.
Similar terms for excitons may be safely neglected due to the large exciton mass $m_{\mathrm{I}}=m_{\mathrm{D}} \sim 10^4 m_{\mathrm{C}}$.

Diagonalization of the Hamiltonian (\ref{Hamiltonian}) gives three dipolariton branches resulting from the strong coupling between the three initial resonances. They are shown as a function of voltage $V$ on the Figure (2a), while Figure (2b) demonstrates the dispersion of the branches for a given applied bias.
The following parameters were taken to qualitatively reproduce the results of Ref. \cite{Coulson2013}: $E_\mathrm{I}(0) = 1.55$ eV, $E_\mathrm{D}(0) = 1.43$ eV, $\gamma = 0.027 \vert e \vert $, $\beta = 7.2$ $10^{-16}$ eV$^{-1}$, $m_C=10^{-4} m_e$, $\Omega= 6 meV$, $J= 6 meV$.
Here $e$ and $m_e$ stand for electron charge and mass.

Each dipolariton eigenstate is a linear combination of excitonic and photonic components with the generalized Hopfield coefficients:
\begin{equation}
\vert \mathbf{Q},S \rangle_{\rm{DP}} = 
\sum_{\rm{j=I,D,C} }c_{\rm{j}}(Q,V) \vert \mathbf{Q},S \rangle{\rm{j}}.
\end{equation}
Here, the index $\mathrm{j}$ spans over indirect ($\mathrm{I}$), direct ($\mathrm{D}$) exciton and cavity photon ($\mathrm{C}$) states.
$\mathbf{Q}$ and $S$ designate quasimomentum and total angular momentum projection (below denoted as spin for simplicity) on the QWs plane (in units of $\hbar$).
These coefficients may be obtained by exact diagonalization of the Hamiltonian (\ref{Hamiltonian}), but their analytical form is quite cumbersome and we do not present them here.

In the following sections we will be interested in the excitonic part of the dipolaritons ($\mathrm{j=I,D}$), since it is responsible for their interactions.
The quantum states $\vert \mathbf{Q},S \rangle_{\rm{I}}$ and $\vert \mathbf{Q},S \rangle_{\rm{D}}$ represent an IX and a DX in the 1s state with center of mass momentum $\mathbf{Q}$ and spin $S=\pm1$, described by wavefunctions having a common form with decoupled motional and spin parts \cite{Nalitov2013}:
\begin{equation}
\Psi^{\mathbf{Q},S} (\mathbf{r}_e,\mathbf{r}_h) =
\Psi^{\mathbf{Q}} \left( \mathbf{R} \right) \Psi^{\rho}(\rho) \Psi^{z}(z_{e},z_{h}) \chi^S(s_e,j_h), 
\end{equation}
where $\mathbf{R}=(m_e \mathbf{r}_e + m_h \mathbf{r}_h)_{\rm{QW}}/(m_e+m_h)$ is the exciton center of mass projection on the QW plane, 
$\mathbf{\rho}=(\mathbf{r}_e - \mathbf{r}_h)_{\rm{QW}}$ is the in-plane distance between the electron and the hole bound into exciton, $z_{e(h)}$ is the electron (hole) coordinate in the QW growth direction and $s_e$, $j_h$ are the electron spin and heavy hole angular momentum projections on the $z$ axis.
The center-of-mass motion part $\Psi^{\mathbf{Q}} \left( \mathbf{R} \right) = S^{-1/2}\exp \left( - \mathrm{i} \mathbf{QR} \right)$, where $S$ is the normalization area, is the same plane wave for both types of excitons.
The internal motion part $\Psi _{\rho}(\mathbf{\rho})$ reads:
\begin{equation}
\Psi _{\rho}(\mathbf{\rho})={\frac{1}{\sqrt{2\pi b(b+r_{0})}}}\exp \left( \frac{-\sqrt{\rho ^{2}+r_{0}^{2}}+r_{0}}{2b}\right) \nonumber,
\end{equation}
where $b$ and $r_0$ parameters are different for IX and DX.
The out-of-plane part may be set as $\Psi^{z}(z_{e},z_{h})=\delta \left( z_e-Z_e \right) \delta \left( z_h-Z_h \right)$, where $Z_e$ and $Z_h$ are the coordinates of the QWs where the electron and the hole are confined. They coincide for DX and differ in the case of IX.
The spin part $\chi^S(s_e,j_h)$ plays a major role in the calculation of the scattering matrix elements as they drastically depend on the spin configuration of a dipolariton pair.
Only the exciton states with a total spin $S=\pm 1$ are coupled to the photonic mode thus forming dipolaritons. For them we define the spin part as $\chi^{\pm 1}(s_e,j_h) = \delta_{s_e,\mp 1/2} \delta_{j_h,\pm 3/2}$.

\section{Polariton-polariton interactions}

\begin{figure} \label{fig3}
\includegraphics[width=0.99\linewidth]{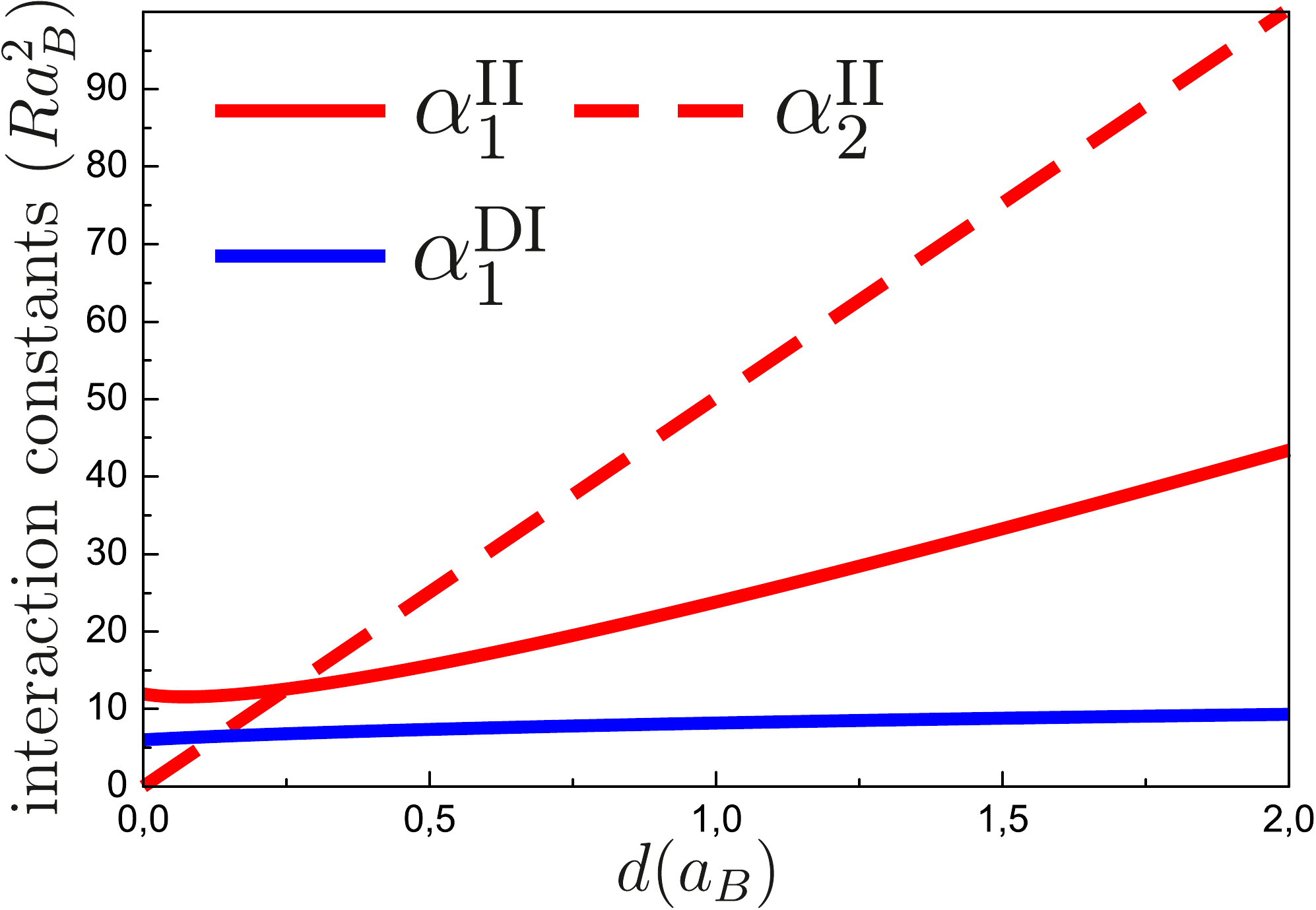}
\caption{Interaction parameters calculated in Born approximation as a function of the separation distance between the QWs. The red lines represent interaction between two IXs, while the blue one corresponds to the interaction of an IX with a DX. The solid lines describe interaction of two excitons with the same spin ($\alpha^{\mathrm{II}}_1$,$\alpha^{\mathrm{DI}}_1$) and the dashed one is for two excitons with opposite spins ($\alpha_2^{\mathrm{II}}$).}
\end{figure}

In this section we shall derive the expressions for the dipolariton-dipolariton scattering matrix elements and then consider the particular case of the parametric scattering of two particles (conserving energy and momentum). We are using the perturbation theory, within the Born approximation in the subsection A. When the corresponding contribution becomes very weak, as it happens for the inter-spin interaction at low applied voltage, we need to proceed further to the second order correction, which is presented in the subsection B.
%
%

\subsection{Born Approximation.}

The wavefunction of a dipolariton pair accounting for the fermionic nature of the carriers is obtained from the direct product of two single dipolariton wavefunctions:
\begin{equation} \label{2DP_WF}
\vert \mathbf{Q},S ; \mathbf{Q}^\prime,S^\prime \rangle =
\left[ \vert \mathbf{Q},S \rangle_{\rm{X}} \otimes \vert \mathbf{Q}^\prime,S^\prime \rangle_{\rm{X}} \right]_a,
\end{equation}
where the index $a$ denotes the antisymmetrization with respect to the  permutations of either electrons ($\mathbf{r}_e \leftrightarrow \mathbf{r}_e^\prime$, $s_e \leftrightarrow s_e^\prime$) or holes ($\mathbf{r}_h \leftrightarrow \mathbf{r}_h^\prime$, $j_h \leftrightarrow j_h^\prime$).

Considering two possible spin configurations of a scattering pair, triplet ($S=S^\prime$) and singlet ($S=-S^\prime$), without loss of generality we write the scattering matrix elements in the Born approximation as:
\begin{equation} \label{V_definition}
V^{(1)}_{f \leftarrow i} = \langle f \vert \hat{V} \vert i \rangle \equiv
\langle \mathbf{Q}_f,S; \mathbf{Q}_f^\prime,S^\prime \vert
\hat{V}
\vert   \mathbf{Q}_i,S; \mathbf{Q}_i^\prime,S^\prime \rangle \notag
\end{equation}
with $\hat{V}$ for the scattering potential, which accounts for the inter-exciton carrier Coulomb interactions:
\begin{equation*}
\hat{V}={\frac{e^{2} }{\epsilon}} \left[ 
\frac{1}{|\mathbf{r}_{e}-\mathbf{r}_{e}^{\prime}|}+
\frac{1}{|\mathbf{r}_{h}-\mathbf{r}_{h}^{\prime}|}- 
\frac{1}{|\mathbf{r}_{e}-\mathbf{r}_{h}^{\prime}|}-
\frac{1}{|\mathbf{r}_{h}-\mathbf{r}_{e}^{\prime }|} \right],
\end{equation*}
where $e$ is the electron charge and $\epsilon$ is the dielectric constant. Without loss of generality, they are decomposed into the following sum:
\begin{align} \label{V_sum}
&V^{(1)}_{f \leftarrow i} = 
\sum_{\rm{i,j,k,l=I,D}} C_{\rm{i,j}}^{\rm{k,l}}(Q_i,Q_i^\prime,Q_f,Q_f^\prime,V) \notag\\ \times
&\left[ \langle \mathbf{Q}_f, S \vert_{\rm{k}} \otimes 
       \langle \mathbf{Q}_f^\prime, S^\prime \vert_{\rm{l}} \right]_a
\hat{V}
\left[ \vert \mathbf{Q}_i, S \rangle_{\rm{i}} \otimes 
       \vert \mathbf{Q}_i^\prime, S^\prime \rangle_{\rm{j}} \right]_a
\end{align}

We consider the range of wave vectors $Q \ll a_B^{-1}$, where $a_B \sim 10 \rm{nm}$ is the bulk exciton Bohr radius.
In this range, all quantum averages in the sum (\ref{V_sum}) are independent on the wave vectors, as their characteristic scale of variation is $a_B^{-1}$.
Therefore, the dependence on the wave vectors as well as on the bias is only kept in the Hopfield coefficients product
$C_{\rm{i,j}}^{\rm{k,l}} =
c_{\rm{i}}  (Q_i,V) c_{\rm{j}}  (Q_i^\prime,V)
c_{\rm{k}}^*(Q_f,V) c_{\rm{i}}^*(Q_f^\prime,V)$.
Finally, the vanishing overlap of DX and IX wavefunctions allows to  keep only the terms where either $\rm{i=k}$, $\rm{j=l}$ or $\rm{i=l}$, $\rm{j=k}$, and to obtain:
\begin{align} \label{Vgeneral}
V^{(1)}_{f \leftarrow i} &= 
C_{\rm{D,D}}^{\rm{D,D}} V_{\rm{D,D}}^{S, S^\prime} +
C_{\rm{I,I}}^{\rm{I,I}} V_{\rm{I,I}}^{S, S^\prime} + \notag\\ &+
\left[ C_{\rm{I,D}}^{\rm{I,D}} + C_{\rm{I,D}}^{\rm{D,I}} +
       C_{\rm{D,I}}^{\rm{I,D}} + C_{\rm{D,I}}^{\rm{D,I}} \right]
V_{\rm{D,I}}^{S, S^\prime}
\end{align}
with interaction constants $V_{\rm{D,D}}^{S, S^\prime}$, $V_{\rm{I,I}}^{S, S^\prime}$ and
$V_{\rm{D,I}}^{S, S^\prime}$ representing DX-DX\cite{Ciuti1998}, IX-IX\cite{Kyriienko2012} and DX-IX\cite{Nalitov2013} interactions. The corresponding matrix elements can be written as follows:
\begin{align} \label{int_const}
V_{\rm{i,j}}^{S, S^\prime} &=
\left[ \langle 0, S \vert_{\rm{i}} \otimes 
       \langle 0, S^\prime \vert_{\rm{j}} \right]_a
\hat{V}
\left[ \vert 0, S \rangle_{\rm{i}} \otimes 
       \vert 0, S^\prime \rangle_{\rm{j}} \right]_a.
\end{align}
where $\mathrm{i,j}$ span over $\mathrm{I,D}$.
The spin dependence of each of the above integrals is then conveniently described by its decomposition into a sum of four terms with evident spin parts:
\begin{equation}
V^{S,S^{\prime}}_{\mathrm{i,j}} = V^{\mathrm{dir}}_{\mathrm{i,j}} +
\delta _{S,S^{\prime }} V^{\mathrm{X}}_{\mathrm{i,j}} +
\delta_{s_{e},s_{e}^{\prime }} V^{\mathrm{e}}_{\mathrm{i,j}} +
\delta_{j_{h},j_{h}^{\prime }}V^{\mathrm{h}}_{\mathrm{i,j}},  \label{V}
\end{equation}

The first term represents the direct dipole-dipole interaction and is present for any combination of exciton spins.
The second term describes the exciton exchange contribution and accounts for the bosonic nature of the exciton.
Finally, the last two terms represent the electron and hole exchange contributions, accounting for the fermionic nature of the carriers.
Interaction constants (\ref{int_const}) are evidently expressed in these terms:
\begin{align} \label{alphas}
\alpha_1^{\mathrm{i,j}} &\equiv V^{+1,+1}_{\mathrm{i,j}} =
V^{\mathrm{dir}}_{\mathrm{i,j}}+
V^{\mathrm{X}}_{\mathrm{i,j}}+
V^{\mathrm{e}}_{\mathrm{i,j}}+
V^{\mathrm{h}}_{\mathrm{i,j}}, \notag \\
\alpha_2^{\mathrm{i,j}} &\equiv V^{+1,-1}_{\mathrm{ij}} =V^{\mathrm{dir}}_{\mathrm{i,j}}
\end{align}

\begin{figure}[h] \label{fig4}
\includegraphics[width=0.99\linewidth]{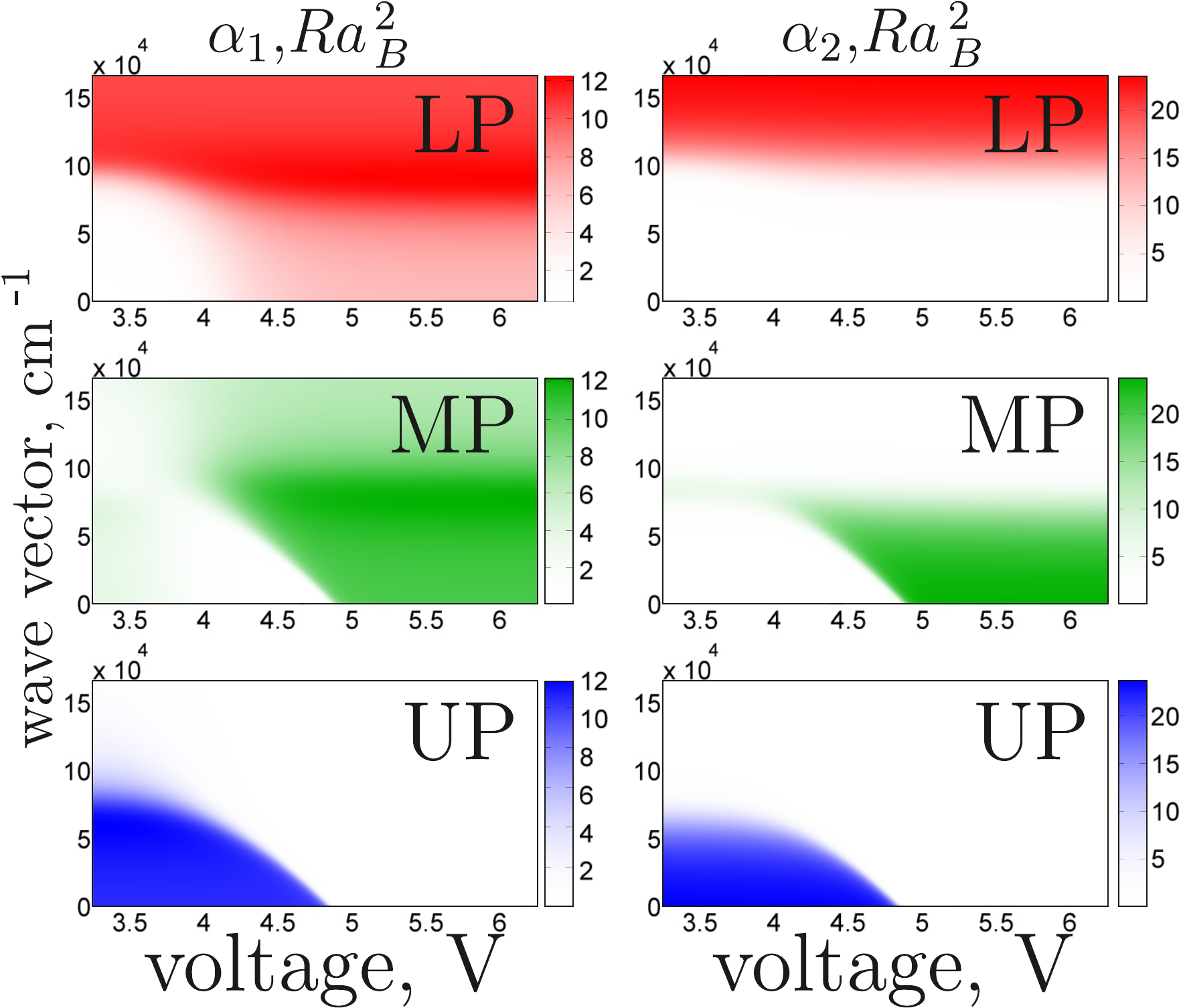}
\caption{ Interaction constants responsible for the blueshift and the bistability of pumped dipolariton states calculated in Born approximation. Left and right columns correspond to $\alpha_1$ and $\alpha_2$, while the three rows correspond to LP, MP and UP dipolariton branches. The voltage and wave-vector dependences are due to the variation of Hopfield coefficients.}
\end{figure}

Neglecting the DX dipole with respect to the one of the IX results in $\alpha^{\mathrm{DD}}_{2} = \alpha^{\mathrm{DI}}_{2}=0$.
Figure 3 shows the dependence of nonzero interaction constants (\ref{alphas}) on the QWs separation distance $d$.
Zero separation limit corresponds to the transition from IX to DX.
At $d \approx a_B/4$, the carrier exchange contribution changes sign and therefore $\alpha^{\mathrm{II}}_2 > \alpha^{\mathrm{II}}_1$ for any reasonable $d>a_B/4 \sim 2\mathrm{nm}$.

Note that $\alpha^{\mathrm{DI}}_1$, plotted by the blue line, is inaccurate in the vicinity of the point $d=0$, where the IX and DX are indistinguishable.
Moreover, the range where the distance between the QWs centers is shorter than their widths is physically meaningless.

Setting $i=f=\vert \mathbf{Q}, S, \mathbf{Q}, S^\prime \rangle$ states at a given point of the energy dispersion branches, we calculate the effective dipolaritonic interaction constants responsible for self-induced blueshifts of pumped dipolaritons luminescence lines:
\begin{align}
\alpha_1 &= \vert c_\mathrm{D}(Q,V) \vert^4 \alpha_1^\mathrm{D,D} + \vert c_\mathrm{I}(Q,V) \vert^4 \alpha_1^\mathrm{I,I} + 4 \vert c_\mathrm{D} \vert^2 \vert c_\mathrm{I} \vert^2 \alpha_1^\mathrm{D,I}, \notag \\ 
\alpha_2 &= \vert c_\mathrm{I}(Q,V) \vert^4 \alpha_2^\mathrm{I,I}.
\end{align} 

Figure 4 presents the results of the numerical calculation of these constants throughout the three dipolaritonic branches in dependence on the applied bias.
Note that $\alpha_{1(2)}$ reflects blueshift of photoluminescence in one circular polarization due to pumping with the same (opposite) circular polarization.
Here $\alpha_{1(2)}$ reflects blueshift of photoluminescence in one circular polarization due to pumping with the same (opposite) circular polarization.

All listed interaction constants are positive, therefore, the dipolariton scattering matrix elements (\ref{int_const})  obtained in this section  only describe repulsive interactions.
In order to include the dipolariton attraction in our model we continue the expansion of the interactions to the second order.

\subsection{Second-Order Corrections}

In this section, we study the second order corrections to the interaction matrix elements, which are important for a scattering of two dipolaritons with opposite spins in the region of voltages and momenta where the IX fraction of either initial or final state are small and so is the 1st order matrix element. This condition can be satisfied for the voltages where $\vert E_{\mathrm{I}}-E_{\mathrm{D}}\vert \gg J$. One can expect the change of sign of the singlet interaction parameter $\alpha_2$ at the point where the 1st and the 2nd order contributions become comparable. 

Scattering of two polaritons with opposite spins has been recently studied theoretically \cite{Glazov2009} and experimentally \cite{Vladimirova2010}.
In particular, the reference \citep{Vladimirova2010} reports strong attraction of singlet lower branch polariton pairs depending on the energy detuning between the DX and Cavity modes.

We start to generalize the results of reference \citep{Glazov2009} to the case of lower branch dipolaritons with the expression for the correction to the scattering matrix element in the second order of the perturbation theory \cite{Landau3}:
\begin{align} \label{SO_interaction}
&V_{f \leftarrow i}^{(2)\uparrow \downarrow} = \sum_{m}\frac{
\langle f \vert \hat{V} \vert m \rangle
\langle m \vert \hat{V} \vert i \rangle }{E_i - E_{m}} \equiv \notag \\
&\equiv \sum_{m}\frac{
\langle \mathbf{Q}_f, +1; \mathbf{Q}_f^\prime, -1 \vert \hat{V} \vert m \rangle
\langle m \vert \hat{V} \vert   \mathbf{Q}_i, +1; \mathbf{Q}_i^\prime,-1 \rangle }{E_i - E_m}
\end{align}
where $m$ enumerates all intermediate states of two electrons and two holes playing the role of the interaction mediator.
%

Intermediate states representing two dipolaritons with opposite spins are coupled with initial and final states by spin-conserving dipole-dipole scattering and result in a second order correction to the repulsion.
On the contrary, states formed by two "dark" excitons
$\vert \mathbf{Q+P},+2 ; \mathbf{Q^\prime-P},-2 \rangle$
are coupled to dipolariton pair states
$\vert \mathbf{Q},+1 ; \mathbf{Q^\prime},-1 \rangle$
via virtual fermion exchange.
Terms with such intermediate states give negative contribution to the interaction potential exceeding the first order repulsion in absolute value.
The same applies to the contribution coming from biexcitonic intermediate states.
It becomes important in the case of polaritonic Feshbach resonance \cite{Takemura2014}, when the dipolariton pair energy coincides with the biexciton energy and expression (\ref{SO_interaction}) diverges.
In this work, we consider polariton states being far from the bi-exciton and dark exciton resonance, so that both terms gives qualitatively the same type contributions. Then, in order to simplify the calculations, we neglect biexciton states contribution and focus on the contribution of dark exciton states.

Due to the electron tunneling through the barrier, the dark IX and DX states are coupled and form two anti-crossing branches $\mathrm{LD}$ and $\mathrm{UD}$ plotted with dashed lines in Figure 2.
Direct and indirect fractions $d^{\mathrm{LD(UD)}}_{\mathrm{D}}$ and
$d^{\mathrm{LD(UD)}}_{\mathrm{I}}$ of the dark branches, obtained by diagonalization of the Hamiltonian (\ref{Hamiltonian}) with $\Omega=0$, are independent of $Q$ due to equal IX and DX effective masses.
Similarly to the previous subsection, we derive the following:
\begin{equation} \label{SO_derived}
V_{f \leftarrow i}^{(2)\uparrow \downarrow} = 
\sum_{\mathrm{i,j,k,l=I,D}}^{\mathrm{m,n=LD,UD}}
C_{\rm{i,j}}^{\rm{k,l}} 
D_{\rm{i,j,k,l}}^{\mathrm{m,n}}
\sum_{\mathbf{P}} \frac{V_{\rm{k,l}}^{\mathrm{exch}}(P) V_{\rm{i,j}}^{\mathrm{exch}}(P)}{-\Delta_{\mathrm{m,n}} - \hbar^2 P^2 / M_X},\notag
\end{equation}
where
$D_{\rm{i,j,k,l}}^{\mathrm{m,n}} (V) = 
d_{\mathrm{i}}^{\mathrm{m}} (V)
d_{\mathrm{j}}^{\mathrm{n}} (V)
d_{\mathrm{k}}^{\mathrm{m}} (V)
d_{\mathrm{l}}^{\mathrm{n}} (V)$ and
$\Delta_{\mathrm{m,n}}(V)=E_{\mathrm{m}}(V)+E_{\mathrm{n}}(V)-E_{i}$.
Here, we neglect once again the dependence of the virtual fermion exchange matrix elements $V^{\mathrm{exch}}_{\mathrm{i,j}}(P)$ on the dipolariton momenta $Q \ll a_B$, although we keep the virtual transferred momentum $\mathbf{P}$ which spans over the whole reciprocal space.
Furthermore, we omit the terms where $\mathrm{i=j=I}$ and $\mathrm{k=l=I}$ representing the next order correction to the IX repulsive contribution.
The virtual exchange matrix elements are:
\begin{align}
V_{\mathrm{I,D}}^{\mathrm{exch}} = 
\left[ \langle 0, +1 \vert_{\rm{I}} \otimes 
       \langle 0, -1 \vert_{\rm{D}} \right]_a
&\hat{V}
\left[ \vert  \mathbf{P}, -2 \rangle_{\rm{I}} \otimes 
       \vert -\mathbf{P}, +2 \rangle_{\rm{D}} \right]_a, \notag \\
V_{\mathrm{D,D}}^{\mathrm{exch}} = 
\left[ \langle 0, +1 \vert_{\rm{D}} \otimes 
       \langle 0, -1 \vert_{\rm{D}} \right]_a
&\hat{V}
\left[ \vert  \mathbf{P}, -2 \rangle_{\rm{D}} \otimes 
       \vert -\mathbf{P}, +2 \rangle_{\rm{D}} \right]_a \notag \\
+\left[ \langle 0, +1 \vert_{\rm{D}} \otimes 
       \langle 0, -1 \vert_{\rm{D}} \right]_a
&\hat{V}
\left[ \vert  \mathbf{P}, +2 \rangle_{\rm{D}} \otimes 
       \vert -\mathbf{P}, -2 \rangle_{\rm{D}} \right]_a. \notag
\end{align}
Note that the two terms of the latter correspond to electron and hole exchange, contrary to the IX-DX case, where only the hole exchange is possible.
Both matrix elements are integrated numerically in dependence on the transferred momentum\citep{Ciuti1998,Nalitov2013} and vanish at $P > a_B/2$.

\begin{figure} \label{fig5}
\includegraphics[width=0.99\linewidth]{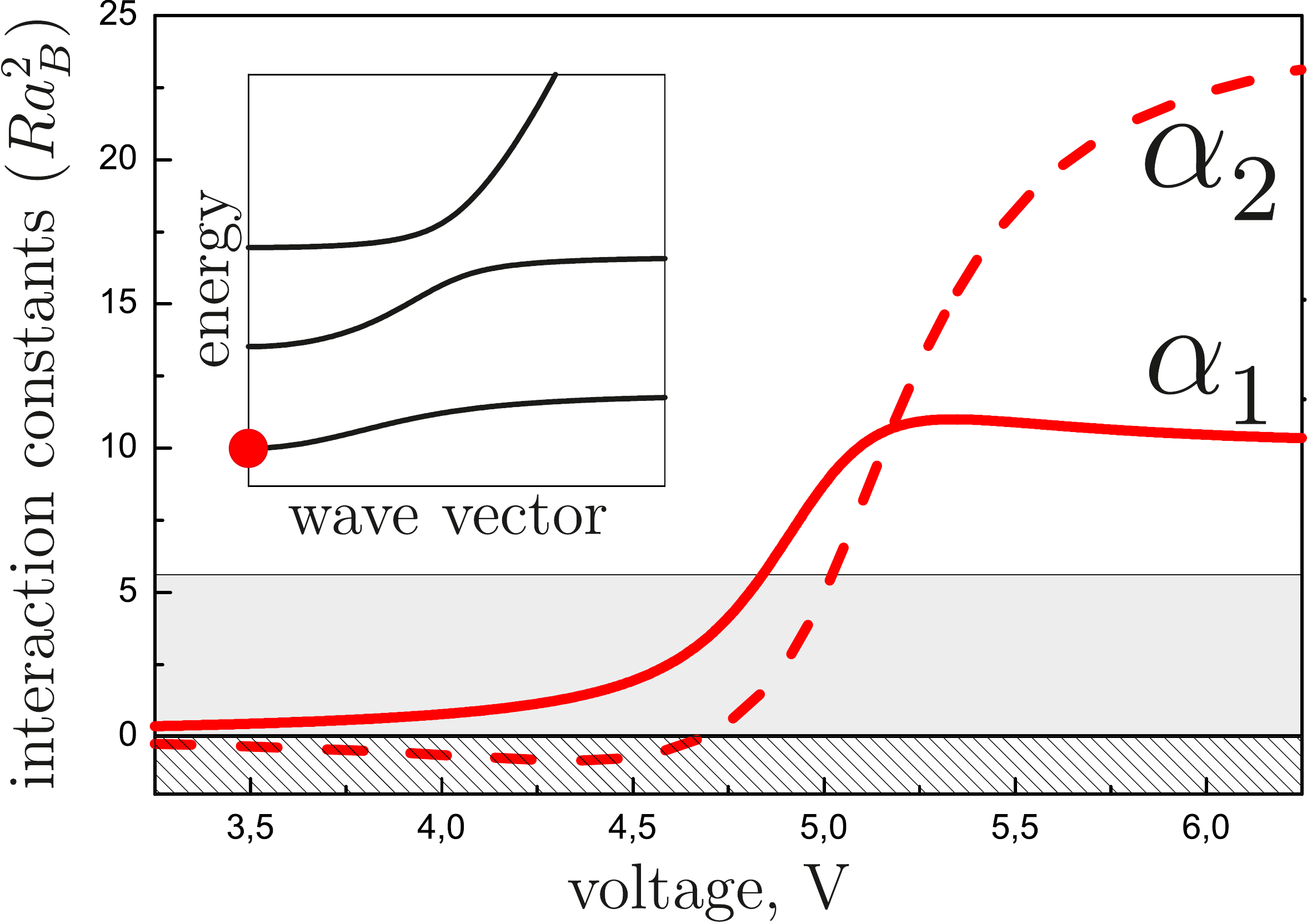}
\caption{Effective interaction parameters responsible for blueshift, bistability and polarization of the ground dipolariton state (sketched in the inset), calculated in the second order of the perturbation theory. $\alpha_2$, describing interaction potential of two dipolaritons with opposite spins, changes sign with votage and even exceeds $\alpha_1$, the one describing two dipolaritons with aligned spins. Grey (hatched) area represent possible values of $\alpha_1$ ($\alpha_2$) for conventional polaritons.}
\end{figure}

Substituting the two-particle ground dipolariton state as both $i$ and $f$ into Eqns. (\ref{Vgeneral},\ref{SO_interaction}) we calculate the effective interaction constants for the ground state.
The results of this calculation is plotted in Figure 5.
One can observe that $\alpha_2$ is changing sign at some particular voltage, similarly with the case of resonant interaction with the bi-exciton resonance \cite{Vladimirova2010,Takemura2014}, which has the disadvantage of inducing strong losses. Here, the mechanism is the increase of the mixing with the IX state which does not add any losses to the dipolarion. However, the whole system can be possibly affected by the large intrinsic losses of the dipolaritonic states, induced by the presence of the metallic contacts and doped mirrors. A second remarkable point occurs at a slightly larger voltage, when $\alpha_2 $ and $\alpha_1$ become equal. In case of a dipolariton condensation, this boundary corresponds to a transition between linearly and circularly polarized states \cite{Vladimirova2010} which can therefore be tuned, simply by changing the applied voltage.

\section{Optical parametric oscillator.}

\begin{figure} \label{fig6}
\includegraphics[width=0.99\linewidth]{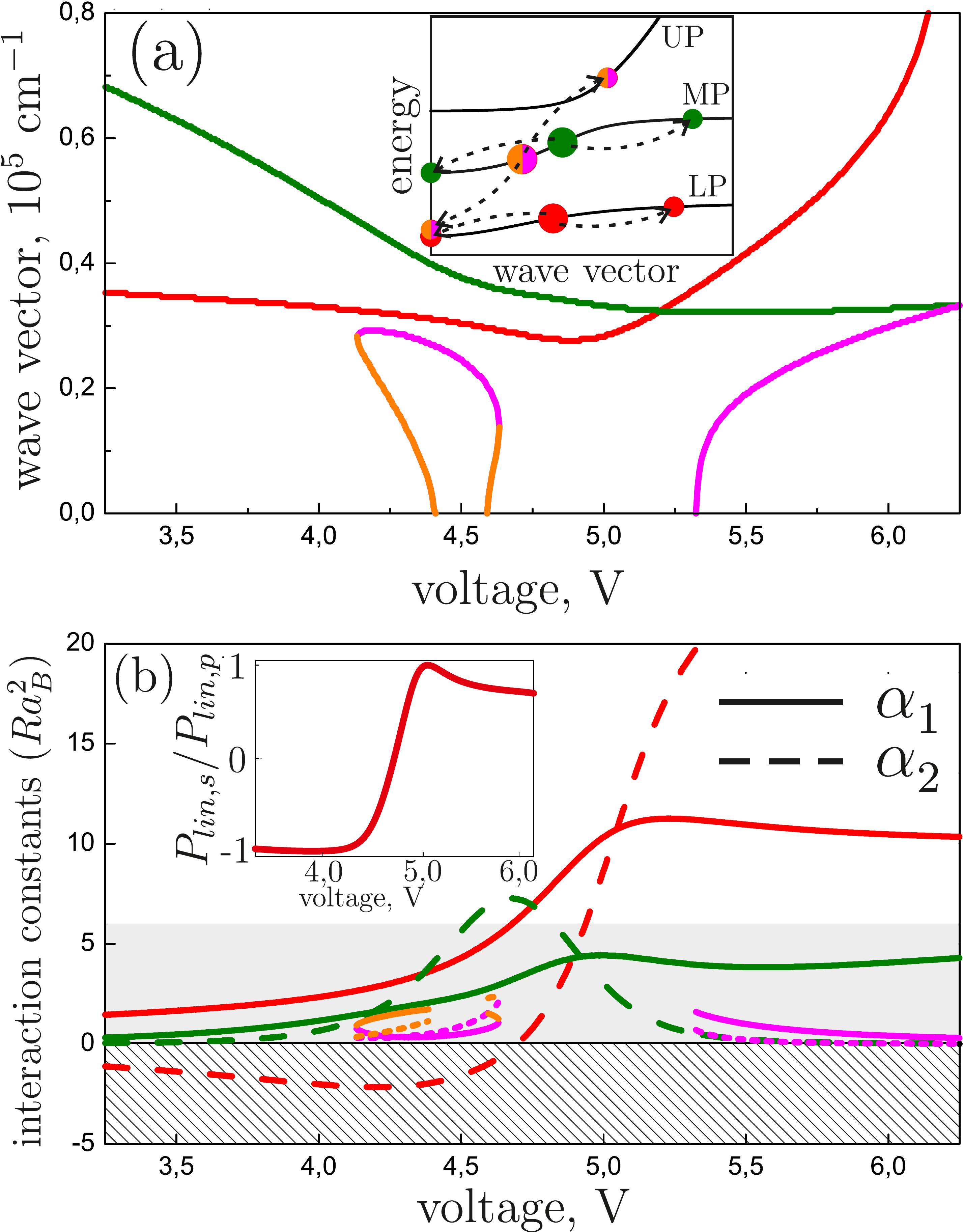}
\caption{Optical parametric oscillator properties. (a) Numerical solution of energy-momentum conservation law for parametric scattering, sketched on the inset. Two dipolaritons in the pump state (large circles) scatter to signal and idler states (small circles). Depending on applied bias there are from 2 to 4 solutions corresponding to different scattering configurations. Wave vector of the pump state, representing the magic angle of optical excitation, is plotted with colours corresponding to configurations: red and green for scattering within LP and MP branches, orange and magenta for interbranch scattering. (b) Interaction constants calculated for dipolariton parametric scattering in the second order of the perturbation theory. Solid lines represent interaction between the dipolaritons with the same spin ($\alpha_1$), dashed lines are for dipolaritons with opposite spins ($\alpha_2$). Grey (hatched) area limits the possible values of $\alpha_1$ ($\alpha_2$) for conventional polaritons. Both LP and MP configurations have a range of voltages where $\alpha_2>\alpha_1$. In the LP case, $\alpha_2$ is  changing sign due to variation of the energy detuning and dipolariton oscillator strength. The inset shows the relation between linear polarization degrees of signal emission and optical pumping in the most relevant case of LP parametric scattering.}
\end{figure}

The OPO configuration of the dipolariton excitation scheme implies parametric scattering of two quasi-particles from the resonantly excited pump state ($Q_p$) conserving the energy and the momentum into the signal ($Q_s=0$) and idler ($Q_i$) states, schematically shown in the inset of Figure (6a).
Numerical solution of in-plane momentum and energy conservation equations
\begin{equation}
2 Q_p = Q_s + Q_i, \quad 2 E_p = E_s + E_i,
\end{equation}
gives all possible pump state wave vectors, satisfying the OPO condition, as a function of the applied voltage, plotted in Figure (6a)
Note, that contrarily to the single-branch OPO schemes, where all states involved in the OPO lie on the same branch (for example, lower polariton branch), the ASDQW structure permits in principle an interbranch OPO scheme with LP signal, MP pump and UP idler states. Such an OPO configuration may be used for generation of entagled photon pairs as both signal and idler states are photonic, therefore the problem of idler polariton coherence loss due to the strong excitonic interactions is avoided.
The UP branch alone does not provide a possibility of a single-branch OPO scheme due to the absence of an inflection point. Moreover, the states lying above the bare exciton energy are resonantly coupled to a large density of excitonic states and can suffer from a significant dephasing, even if their excitonic fraction is small. This dephasing is not accounted for in our approach. The resonant pumping of the LP is therefore the only configuration which we expect to be properly described by our approach.

Spin kinetics of the system strongly depends on the type of polariton-polariton interaction.
Polarization of photons, emitted from the signal state once the OPO turns on, is defined by the one of pumping and two interaction constants $\alpha_1$ and $\alpha_2$ describing parametric scattering of a pair of dipolaritons with aligned and anti-aligned spins respectively.
In the particular case of linearly polarized pumping, the signal linear polarization degree is expressed by a simple relation \cite{Renucci2005}:
\begin{equation} \label{pol}
P_{\textit{lin},s} = \frac{\alpha_1 \alpha_2}{\alpha_1^2+\alpha_2^2} P_{\textit{lin},p},
\end{equation}
where $P_{\textit{lin},p}$ is the linear polarization of the optical pumping.
The sign of $P_{\textit{lin},s}$ and thus the orientation of the signal polarization plane are therefore determined by the relative sign of the interaction parameters $\alpha_1$ and $\alpha_2$ describing parametric scattering of dipolaritons with aligned and opposite spins. 

To calculate them accounting for the second order correction we substitute  the pump, signal and idler dipolariton states into Eqs.(\ref{Vgeneral},\ref{SO_interaction}):
\begin{align}
\alpha_1 &= c_\mathrm{D}^2(Q_p,V)c_\mathrm{D}(Q_s,V)^* c_\mathrm{D}(Q_i,V)^* \alpha_1^\mathrm{D,D} \notag \\
&+c_\mathrm{I}^2(Q_p,V)c_\mathrm{I}(Q_s,V)^* c_\mathrm{I}(Q_i,V)^* \alpha_1^\mathrm{I,I} \notag \\
&+ 2 c_\mathrm{I}(Q_p,V) c_\mathrm{D}(Q_p,V))[ c_\mathrm{D}(Q_s,V)^* c_\mathrm{I}(Q_i,V)^* \notag \\ 
&+ c_\mathrm{D}(Q_s,V)^* c_\mathrm{I}(Q_i,V)^* ] \alpha_1^\mathrm{D,I}, \notag \\
\alpha_2 &= c_\mathrm{I}^2(Q_p,V)c_\mathrm{I}(Q_s,V)^* c_\mathrm{I}(Q_i,V)^* \alpha_2^\mathrm{I,I} \notag \\
+ & \sum_{\mathrm{i,j,k,l=I,D}}^{\mathrm{m,n=LD,UD}}
c_\mathrm{i}(Q_p,V)c_\mathrm{j}(Q_p,V)c_\mathrm{k}(Q_s,V)^*c_\mathrm{l}(C_i,V)^* \notag \\
&D_{\rm{i,j,k,l}}^{\mathrm{m,n}}
\sum_{\mathbf{P}} \frac{V_{\rm{k,l}}^{\mathrm{exch}}(P) V_{\rm{i,j}}^{\mathrm{exch}}(P)}{-\Delta_{\mathrm{m,n}} - \hbar^2 P^2 / M_X},
\end{align}

The results of this calculation are plotted in Figure (5b) for LP, MP and interbranch scattering configurations.
Substituting the pump, signal and idler dipolariton states into $i$ and $f$ we finally calculate the interaction constants
\begin{align}
\alpha_1 &= \langle Q_s,+1;Q_i,+1 \vert \hat{V} \vert Q_p,+1,Q_p,+1 \rangle
\notag \\
\alpha_2 &= \langle Q_s,+1;Q_i,-1 \vert \hat{V} \vert Q_p,+1,Q_p,-1 \rangle +
V^{(2)\uparrow\downarrow}_{f \leftarrow i} \notag
\end{align}
for LP, MP and interbranch OPO configurations, plotted in Figure (5b).
Notably, the following situations may be achieved by voltage variation for different OPO configurations: (i) $\alpha_2<0$, linear polarization inversion is  on; (ii) $0<\alpha_2<\alpha_1$ and (iii) $\alpha_2>\alpha_1$, linear polarization inversion is off.
Moreover, as can be seen in Figure (5b), in a certain range of voltages, the dipolaritonic OPO interaction constant exceeds the theoretically achievable value of $\alpha_1 = 6 R a_B^2$ for conventional polaritons \citep{Glazov2009}.

Substitution of the calculated interaction constants into relation \ref{pol} finally gives the dependence of the signal linear polarization degree on the applied bias for the case of full linear polarization of the pumping ($P_{\textit{lin},p}$), plotted in the inset of Figure (5b).
It has a fast switching region from negative to positive values in the vicinity of the crossing point of the three modes, where $\alpha_2$ value crosses zero.
Realistically, the absolute value of the signal polarization degree is lowered by spin relaxation processes.
However, the main result is that the orientation of the signal emission polarization plane may be switched between the one of the optical pump and the one orthogonal to it.

\section{CONCLUSION}
We have calculated the spin-dependent interaction parameters for dipolaritons and analyzed the specific role played by the dipolar interaction between indirect excitons. We have shown that these parameters can be one order of magnitude larger than for conventional polaritons. By tuning the applied voltage, the interaction parameter $\alpha_2$ between dipolaritons with opposite spin changes sign and can become larger than $\alpha_1$ -- the interaction parameter between dipolaritons having the same spin. 

This shows that dipolaritons are promising particles for spin-optronic applications. In particular, we consider a dipolaritonic OPO scheme, which, due to the large values of the interaction parameters, has a very low threshold. It offers the possibilities of interbranch parametric scattering. The flipping of the sign of the singlet interaction $\alpha_2$ parameter allows the on-demand linear polarization inversion switching and polarization degree control by the applied bias. 

This work has been supported by EU FP7 ITN INDEX (289968) and ANR QUANDYDE (ANR-11-BS10-001). We thank M. Glazov for fruitful discussions.

\bibliography{reference}

\end{document}